\documentclass[11pt,a4paper]{amsart}

\usepackage{graphicx}      
\usepackage{amsmath}
\usepackage{amssymb}
\usepackage{xcolor}
\usepackage{amsthm}
\usepackage{appendix}
\usepackage{epstopdf}
\usepackage{gensymb}
\usepackage{placeins}
\usepackage{algorithm}
\usepackage{graphicx}
\usepackage{microtype}
\newcommand{\mrm}[1]{\mathrm{#1}}

\usepackage{tikz}
\usetikzlibrary{shadows,arrows,positioning}
\usepackage{algpseudocode}

\theoremstyle{definition}

\begin{document}
\title{Model predictive control for retinal laser treatment at 1\,kHz}
\author[M.\ Schaller et al.]{Manuel Schaller$^{1}$, Viktoria Kleyman$^{2}$, Mario~Mordmüller$^{3}$, Christian~Schmidt$^{3}$, Mitsuru Wilson$^{1}$, Ralf~Brinkmann$^{3,4}$, Matthias A.\ Müller$^{2}$ and Karl~Worthmann$^{1}$}
\thanks{}

\thanks{$^{1}$Technische Universit\"at Ilmemau, Institute of Mathematics, Germany (e-mail: \{manuel.schaller,mitsuru.wilson,karl.worthmann\}@tu-ilmenau.de).}
\thanks{$^{2}$Leibniz University Hannover, Institute of Automatic Control, Germany (e-mail: \{kleyman,mueller\}@irt.uni-hannover.de).}
\thanks{$^{3}$University of Lübeck, Institute of Biomedical Optics, Germany (e-mail: \{ralf.brinkmann,m.mordmueller\}@uni-luebeck.de).}

	\thanks{{\bf Acknowledgments: }	The collaborative project "Temperature controlled retinal laser treatment" is funded by the German Research Foundation (DFG) under the project number 430154635 (MU~3929/3-1, WO~2056/7-1, BR~1349/6-1). 
		Karl Worthmann gratefully acknowledges funding by the German Research Foundation (DFG; grant WO~2056/6-1, project number 406141926).}

\begin{abstract}
Laser photocoagulation is a technique applied in the treatment of retinal diseases. While this is often done manually or using simple control schemes, we pursue an optimization-based approach, namely Model Predictive Control (MPC), to enforce bounds on the peak temperature and, thus, safety during the medical treatment procedure --~despite the spot-dependent absorption of the tissue. 
To this end, a repetition rate of $1$\,kHz is desirable rendering the real-time requirements a major challenge. We present a tailored MPC scheme using parametric model reduction, an extended Kalman filter for the parameter and state estimation, and suitably constructed stage costs
and verify its applicability both in simulation and experiments with porcine eyes. Moreover, we give some insight on the implementation specifically tailored for fast numerical computations. 

\smallskip
\noindent \textbf{Keywords.}      Retinal photocoagulation, model predictive control, real-time control
\end{abstract}

\maketitle
\section{Introduction} 
\label{sec:intro}
Laser photocoagulation is a treatment for a variety of retinal diseases. Until now, the physician has to manually adjust the laser power from treatment spot to treatment spot. As the absorption in the eye is highly spot-dependent, the information from previous spots is used to determine the dosage. This might not be sufficient in order to ensure both effectivity and safety of the treatment, that is, to guarantee a sufficient temperature increase without taking the risk of overtreatment.

\begin{figure}[h]
	\centering
	\includegraphics[width =.7\linewidth]{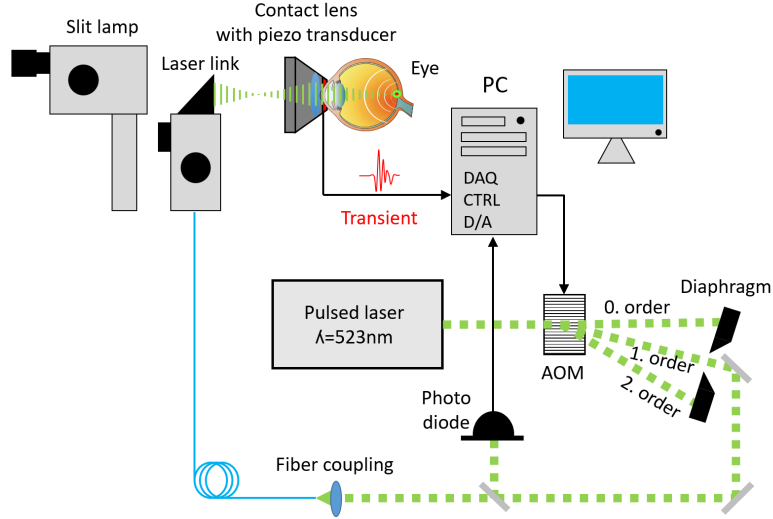}
	\caption{Schematic sketch of the experimental setup, adapted from~\cite{Kleyman2021}}
	\label{fig:Setup_01}
\end{figure}

In order to enable non-invasive feedback control techniques, a possible route is to measure pressure transients resulting from a temperature increase in the eye and to compute an averaged depth-weighted volume temperature as suggested in \cite{Brinkmann.2012}. After focusing a laser beam corresponding to a certain power onto the eye fundus, these pressure transients are measured with a ring-shaped piezo-electric transducer inside the contact lens, as depicted in Fig.~\ref{fig:Setup_01}.
This pressure transient signal is then processed by means of C/C++ data acquisition software and the corresponding volume temperature is computed. Together with an appropriate model of the volume temperature by means of an output of the system, we are able to design feedback control techniques that actuate the laser power that is applied in the next control loop. Again, the resulting pressure transients are measured and the process is repeated. For further details considering the experimental setup, we refer to our previous work  \cite{Mordmueller2021}.

\noindent One key aspect in this medical application is that the peak temperature, that is, the maximal temperature in the eye fundus, plays a predominant role w.r.t.\ coagulation of the tissue. As this peak temperature is not measurable by means of non-invasive methods, we propose a model-based approach using Model Predictive Control (MPC; \cite{Gruene2016b,rawlings2017model}) in order to control this peak temperature to ensure a safe and effective treatment. This approach is different to previous control strategies in \cite{Baade.2017,Herzog.2018}. In these works, a constant relation between the absorption coefficients is assumed to allow for an offline approximation of a direct conversion between the measured volume temperature and the corresponding control target, i.e., the peak temperature. Based on this conversion between both temperatures an open-loop control was designed in \cite{Baade.2017} and a closed-loop control in \cite{Herzog.2018}. However, we found in our work \cite{Schaller2022c}, that the assumption of a constant relation between both absorption coefficient does not hold in general. 

Hence, in our approach presented here, we model the underlying absorption and heat diffusion process in state space in order to enable peak temperature control as one system-theoretic output of the system, while measuring the volume temperature as a second output and without considering a direct conversion. The absorption of the laser beam and the heat diffusion is governed by a parametric partial differential equation (PDE) that depends on the (highly) patient and spot specific absorption coefficient. Hence, after spatial discretization of this PDE, we obtain a high-dimensional parametric control system \cite{Kleyman2020b}, where the state is a discrete version of the temperature distribution in the eye fundus.

In order to make this model accessible for real-time MPC, several components that we addressed in previous publications are crucial. In \cite{Schaller2022c}, we tailored and compared methods for parametric model reduction in the context of our application. Further, as the model depends on the unknown state and absorption parameter of the tissue, we consider in \cite{Kleyman2021,Kleyman2022} an extended Kalman filter~(EKF) approach for joint state/parameter estimation.

In \cite{Mordmueller2021}, we recently presented preliminary MPC closed-loop results at 250\,Hz in experiments, using a reduced order model based on Taylor-approximation of the parametric dependency as suggested earlier in \cite{Kleyman2020b}. Regarding the MPC stage-cost design, we considered quadratic penalization of the control size of the form $|u|^2$. Here, we will consider parametric model reduction approach which recently in \cite{Schaller2022c} was found to be superior and that is based on Discrete Empirical Interpolation (\cite{Chaturantabut2010}; DEIM). Moreover, we will present and evaluate different choices of the stage cost, including a non-autonomous state cost, a control penalization by means of $|u-u_\text{ref}|^2$ with an adaptively chosen reference $u_\text{ref}$, and a penalization for the control variation. Last, we will present also results for a higher sampling rate of 1\,kHz.

\begin{figure}[h]
	\centering
\pgfdeclarelayer{background}
\pgfsetlayers{background,main}

\tikzstyle{materia}=[draw, fill=gray!15, text width=8.0em, text centered,
minimum height=1.5em,drop shadow]
\tikzstyle{etape} = [materia, minimum width=5em,
minimum height=3em, rounded corners, drop shadow]
\tikzstyle{line} = [draw, thick, color=black!80, -latex']
\tikzstyle{helper} = [draw, shape=circle, inner sep = 0pt, minimum size = .005cm, black]

\newcommand{\blockdist}{1.3}
\newcommand{\edgedist}{1.5}

\newcommand{\etape}[2]{node (p#1) [etape]
	{#2}}

\newcommand{\background}[5]{%
	\begin{pgfonlayer}{background}
		\path (#1.west |- #2.north)+(-0.5,0.25) node (a1) {};
		\path (#3.east |- #4.south)+(+0.5,-0.25) node (a2) {};
\end{pgfonlayer}}

\begin{tikzpicture}[scale=0.9,transform shape]
	\path \etape{1}{Initialize state and absorption coefficient};
	
	\path (p1.south)+(0.0,-2) \etape{2}{MPC: Solve optimal control problem to obtain input $u$};
	\path (p2.south)+(0.0,-2) \etape{3}{EKF: Estimate state $x$ and parameter $\alpha$};
	
	\path (p2.east)+(2.8,0) \etape{4}{Setup};
	\node[helper] (h1) [left of = p3, xshift=-2cm] {};
	
	\path [line] (p1.south) -- node [left] {$x^0,\,\alpha^0$} (p2);
	\path [line] (p2.south) -- node [left] {$u$} (p3);
	\path [line] (p2.east) -- node [above] {$u$} (p4);
	\path [line] (p4.south) |- node [above,xshift=-1.4cm] {measurement} (p3);
	\draw[-,thick] (p3.west) -- (h1);
	\path [line] (h1) |- node [right, yshift=-1.5cm] {$x,\,\alpha$} (p2);
	
\end{tikzpicture}
	\caption{Model-based control scheme for retinal laser treatment}
	\label{fig:flowchart}
\end{figure}
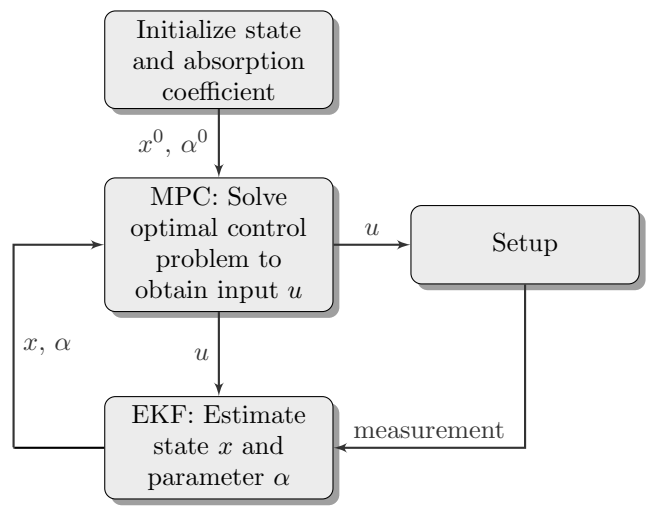

The main contribution of this work is the combination of the above ingredients with MPC in real-time experiments. We briefly summarize our approach in Fig.~\ref{fig:flowchart}. We use a parametric reduced order model (pROM) obtained in \cite{Schaller2022c} in the EKF to allow for joint state and parameter estimation as presented in \cite{Kleyman2022}, which subsequently enables the computation of a laser power by means of an MPC feedback based on the current estimate of absorption.

\noindent This paper is organized as follows. In Section~\ref{sec:modeling} we introduce the underlying partial differential equation describing the absorption and heat diffusion and present the output operators corresponding to the volume and the peak temperature. Further, we briefly describe the corresponding discretization, the parametric model reduction and the joint state/parameter estimation using an extended Kalman Filter that are used in the following. Then, in Section~\ref{sec:MPC}, we present a  particularly tailored optimal control problem that is used to compute the MPC feedback. In Section~\ref{sec:250} we present MPC closed-loop results both in simulation and in experiments using porcine eyes at a repetition rate of 250\,Hz. Subsequently, we provide results of closed-loop control in 1\,kHz showing that that a higher frequency is desirable. Last, in Section~\ref{sec:conc} we conclude by means of an outlook with respect to future work.

\section{Modeling and joint state/parameter estimation} 
\label{sec:modeling}
We briefly describe the PDE model, its discretization and the subsequent model reduction, as well as the state and parameter estimation technique we pursue. For more details, the reader is referred to \cite{Kleyman2020b,Kleyman2021}.

\begin{figure}[h]
	\centering  
		\includegraphics[width =.6\linewidth]{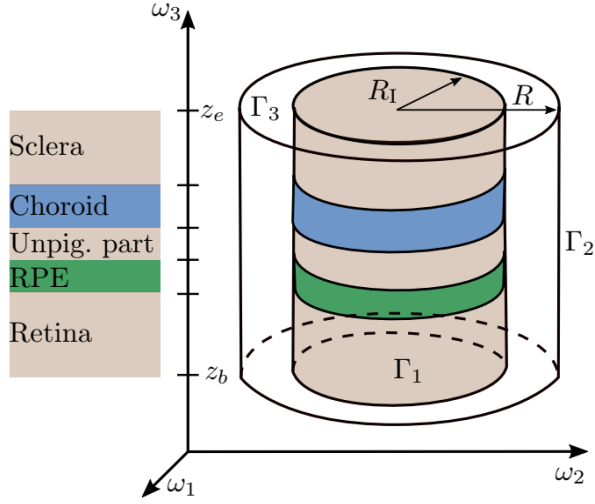}
	\caption{Schematic illustration of computational domain. Fig. adapted from \cite{Kleyman2020b}.}
	\label{fig:cylinders}
\end{figure}

The computational domain in the eye fundus, denoted by $\Omega \subset \mathbb{R}^3$, is depicted in Fig.~\ref{fig:cylinders}. We consider a cylinder with radius $R$ that encloses a smaller cylinder with radius $R_\text{I}$ describing the irradiated area. The absorption in the five different layers is modeled by a piecewise constant function $\mu:\mathbb{R}^+ \to \mathbb{R}^+$ which, due to radial symmetry, only depends on the depth variable $\omega_3$.
\medskip

\noindent Following the Lambert-Beer law describing the absorption of the laser light in the tissue, the evolution of the temperature distribution $x(t,\omega)$ at time instance $t \geq 0$ and spatial coordinate $\omega \in \mathbb{R}^3$ can be described via the parabolic partial differential equation (PDE) of heat diffusion
\begin{align*}
\rho C_\text{p} \frac{\partial x(t,\omega)}{\partial t}-k\Delta x(t,\omega)&=u(t)\frac{\chi_{\text{I}}(\omega)}{\pi R_\text{I}^2}\mu(\omega_3) e^{-\int_0^{\omega_3}\!\!\mu(\zeta)\text{d}\zeta}\quad &&\forall(t,\omega)\in \mathbb{R}^+\times \Omega
\\
x(t,\omega)&=0&&\text{$\forall\, (t,\omega)\in\mathbb{R}^+\times \Gamma$},\\ 
x(0,\omega)&=0 &&\forall\, \omega\in\Omega,
\end{align*}
where $\Delta = \frac{\partial^2}{\partial ^2 \omega_1}+\frac{\partial^2}{\partial ^2 \omega_2}+\frac{\partial^2}{\partial ^2 \omega_3}$ is the usual Laplace operator, $\Gamma~=~\Gamma_1\cup\Gamma_2\cup\Gamma_3$ is the spatial boundary of the domain $\Omega$, cf.\ Fig.~\ref{fig:cylinders}, and $\chi_{\text{I}}$ is the characteristic function of the inner cylinder where the absorption takes place. Here, ${\rho = 993\text{ kg/m}^3}$ is the density, $C_\text{p} = 4176\text{ J/(kgK)}$ the heat capacity and ${k = 0.627\text{ W/mK}}$) the thermal conductivity of water, which composes the main component of the tissue, cf.~\cite{Baade.2017}.

Due to linearity of this equation in the state, we consider temperature increases relative to the ambient temperature and hence impose a homogeneous initial condition. Further, the outer cylinder in Fig.~\ref{fig:cylinders} is chosen large enough such that we can safely assume homogeneous Dirichlet boundary conditions.

The volume temperature that can directly be computed from the measured pressure transients, cf.\ Fig.~\ref{fig:Setup_01}, and the peak temperature that is central for control are defined via
\begin{align*}
\mathcal{C}_\text{vol}(\mu)x &= \int_{z_\text{b}}^{z_\text{e}} x_\text{mean}(t,\omega_3) \mu(\omega_3)e^{\int_{0}^{\omega_3}\mu(\zeta)\text{d} \zeta}\, \text{d}\omega_3,\\
\mathcal{C}_\text{peak}x &= x(0,0,z_\text{center}),
\end{align*}
where $x_\text{mean}(\omega_3)$ is the mean over the radial variable at depth $\omega_3$ and $z_\text{center}$ is the midpoint of the RPE layer.
The absorption coefficient is defined piecewise for each layer in Fig.~\ref{fig:cylinders}, i.e.,
\begin{align*}
\mu(\omega_3) = \begin{cases}
\mu_\text{RPE}, \quad &\text{if } \omega_3 \in \text{RPE},\\
\mu_\text{ch}, \quad &\text{if } \omega_3 \in \text{choroid},\\
0, &\text{otherwise}.
\end{cases}
\end{align*}
We consider the absorption coefficient in the choroid $\mu_\text{ch} = 2662.2$\, m$^{-1}$ to be identical for each spot, cf.\ \cite{Schaller2022c} for a thorough case study. Further, we parameterize the absorption in the RPE with a scalar prefactor $\alpha > 0$, i.e.,
\begin{align*}
\mu_\text{RPE}(\alpha) = \alpha \mu_\text{RPE}^0,
\end{align*}
where $\mu_\text{RPE}^0 = 120400$\,m$^{-1}$ is a reference value from the literature \cite{Brinkmann.2012}. Thus, from now on we will denote dependencies on the absorption coefficient $\mu$ using its parameterization $\alpha$.

Exploiting radial symmetry of the computational domain and applying finite differences in space, cf.\ our previous work \cite{Kleyman2020b}, yields the high-dimensional state-space model
\begin{align}
\nonumber
\dot{x}(t)&= A_\text{f}x(t) + B_\text{f}(\alpha)u(t)\\
\label{eq:fullmodel}
y_\text{vol}(t)&=C_\text{f,vol}(\alpha)x(t)\\
\nonumber
y_\text{peak}(t)&=C_\text{f,peak}x(t)
\end{align}
with $n\in \mathbb{N}$ large (>80000), $A_\text{f}\in \mathbb{R}^{n\times n}$ and $B_\text{f},C_\text{f,vol}:\mathbb{R}\to \mathbb{R}^n$ and $C_\text{f,peak}\in \mathbb{R}^n$. 

After parametric Model Order Reduction by means of a global basis approach \cite[Section 4.1]{benner2015survey} and discrete empirical interpolation \cite{Chaturantabut2010} that we tailored to our application in \cite{Schaller2022c}, and implicit Euler discretization with suitable stepsize, we obtain the low-dimensional parametric surrogate model
\begin{align}
\nonumber
x_{k+1}&= Ax_k + B(\alpha)u_k\\
\label{eq:redmodel}
y_{\text{vol},k}&=C_\text{vol}(\alpha)x_k\\
\nonumber
y_{\text{peak},k}&=C_\text{peak}x_k.
\end{align}
for $k\in \mathbb{N}_0$, where all matrices are reduced to the state space dimension six.
As step sizes for time discretization we will either choose 0.004\,ms or 0.001\,ms corresponding to the repetition rates of 250\,Hz and 1\,kHz that we will consider in Sections~\ref{sec:250} and \ref{sec:1}, respectively.

As a last component in order to enable model-based control via MPC, we have to estimate the unknown spot-dependent absorption coefficient $\alpha$ and the current state $x$, i.e., the temperature distribution, of the system. To this end, we employ an extended Kalman filter (EKF) as a well known state estimator for nonlinear systems using a successive linearization of a nonlinear model that is subject to uncorrelated and normally distributed process noise and measurement noise at each time step , see e.g. \cite{Chui2017}. In our application, we utilize the EKF both for state and parameter estimation by extending the state variable $\bar{x} = (x,\alpha)$ and considering constant parameter dynamics, that is, we apply the EKF to the nonlinear extended dynamics
\begin{align*}
\begin{split}
\bar{x}_{k+1} &= 
\begin{pmatrix}
x_{k+1} \\ \alpha_{k+1} \end{pmatrix} = \begin{pmatrix}
A & 0\\ 0 &1 
\end{pmatrix} \bar{x}_k + \begin{pmatrix}
B(\alpha_k)\\ 0
\end{pmatrix} u_k, \\ 
y_k &= \begin{pmatrix}
C_{\mrm{vol}}(\alpha_k) & 0
\end{pmatrix} \bar{x}_k
\end{split}.
\end{align*}
For more details regarding parameter and state estimation, we refer to our work \cite{Kleyman2021,Kleyman2022}.

\section{Model predictive control}
\label{sec:MPC}
After presenting all necessary ingredients to enable model-based control, we now present the MPC controller as a last component of our scheme, cf.\ Fig.~\ref{fig:flowchart}.
Here, we aim to increase the peak temperature to a reference peak temperature $y_\text{peak,ref}$ (effectivity of the treatment) without exceeding a maximal peak temperature $y_\text{peak,max}$ (safety of the treatment) by means of MPC using a maximal laser power $u_\text{max}=0.1$\,W. We denote by $n\in \mathbb{N}_0$ the iteration counter of the control loop depicted in Fig.~\ref{fig:flowchart}. For the $n$-th MPC step, we assume that we have an estimate for the absorption coefficient $\alpha^n$ and the current state $x^n$ by means of EKF as described in the previous section. Then, for a prediction horizon $N\in \mathbb{N}$, $N \geq 2$ we consider the following optimal control problem:
\begin{alignat}{2}
\nonumber
\min_{u\in \mathbb{R}^{N-1}}&J_n(x,u)\\
\nonumber
\text{s.t. } x_{k+1} &= A x_k + B(\alpha^n)u_k \,\,\,&&k=0,\ldots,N-2\\
\label{eq:OCP}\tag{OCP($n$)}
x_0 &= x^n\\
\nonumber
0&\leq u_k \leq u_\text{max} && k=0,\ldots, N-2\\
\nonumber
C_\text{peak}x_k &\leq y_\text{peak,max} && k=0,\ldots, N-1.
\end{alignat}
The cost functional consists of several terms that are scaled via a possibly time-varying state cost penalization $R_0:\mathbb{N}\to\{0,1\}$ and two scalar control cost parameters $R_1,R_2 > 0$, i.e., 
\begin{align}
\label{eq:cost}
\begin{aligned}
J_n(x,u) &= \sum_{k=0}^{N-1}  R_0(n+k)|C_\text{peak}x_k -y_\text{peak,ref}|^2 \\&+ R_1|u_k-u_{\text{ref},n}|^2 \\&+ R_2\left(\sum_{k=1}^{N-1}|u_k-u_{k-1}|^2 + |u_0 - u_{\text{prev},n}|^2\right).
\end{aligned}
\end{align}
The value $u_{\text{ref},n}$ is the steady-state control corresponding to the steady-state output $y_\text{peak,ref}$ computed by means of system inversion with the current estimate of the absorption coefficient $\alpha^n$, i.e.,
\begin{align*}
u_{\text{ref},n} = (C_\text{peak}A^{-1}B(\alpha^n))^{-1}y_\text{ref}.
\end{align*}
Further, we denote by $u_\text{prev}$ the MPC feedback, that is, the first optimal control value $u_0$ of the preceding MPC iteration. For the first MPC iteration, i.e., $n=0$, we set $\alpha^0 =0.7363$ and $u_{\text{prev},0}=u_{\text{ref},0}$, the former being the empirical mean of the absorption coefficient in porcine eyes obtained from a thorough case study in \cite{Schaller2022c}.
A time-varying penalization $R_0$ that sets the state cost to zero for the first few time steps in order to prevent aggressive controls will be defined and compared to the standard case $R_0\equiv 1$ in Section~\ref{sec:250}. Further, whereas the parameter $R_1$ directly penalizes the control size, resp.\ its deviation from the reference value $u_{\text{ref},n}$, the parameter $R_2$ can be understood as a penalization of variations in the control signal in order to smoothen the control signal to not overreact on inevitable errors in state and parameter estimation.

In this work, we inspect the MPC closed-loop performance without a prior phase that is solely used for identification of the absorption coefficient, which for the sake of safety will almost surely be present in the final medical product. Here, however, this choice is motivated in order to evaluate MPC under demanding conditions as, at the beginning of the treatment, we might still have possibly large errors in the parameter estimation. An underestimation of the actual absorption parameter results in a high laser power obtained as MPC feedback, which, when applied to the eye fundus enjoying a higher absorption than expected might result in overshooting both the targeted peak temperature and the peak temperature bounds. To prevent this, we present a suitable scaling of the terms in the cost functional, i.e., $R_0$, $R_1$ and $R_2$ in \eqref{eq:cost} both in simulation and in experiments in Sec.~\ref{sec:250}.

The scheme in Fig.~\ref{fig:flowchart} can now be specified in Algorithm~\ref{alg:mpc}, where the initialization of the absorption coefficient $\bar{\alpha} = 0.7363$ is the empirical mean.
\begin{algorithm}
	\caption{Model-based retinal laser treatment.}\label{alg:mpc}
	\begin{algorithmic}[1]
		\State Initialize $x\gets 0$, $\alpha \gets \bar{\alpha}$, $n\gets 0$
		\While{Treatment}:
		\State Solve \eqref{eq:OCP} s.t.\ init.\ state $x^n$ and param.\ $\alpha^n$ \label{algline:ocp}
		\State Apply first optimal control signal $u_0$
		\State Measure resulting volume temperature $y_\text{vol}$
		\State $(x^n,\alpha^n) \gets$ EKF with output $y_\text{vol}$.
		\State $u_{\text{ref},n} \gets (C_\text{peak}A^{-1}B(\alpha^n))^{-1}y_\text{ref}$.
		\State $n\gets n+1$
		\EndWhile
	\end{algorithmic}
\end{algorithm}

\noindent We briefly discuss particularities of the implementation that are crucial to render the software capable of real-time control.\\
\noindent \textbf{Implementation details.}
The optimal control problem in Line~\ref{algline:ocp} of Algorithm~\ref{alg:mpc} is solved using the C++-library OSQP~\cite{osqp}, a highly optimized solver for linear quadratic problems with state and control constraints. 
Its implementation is particularly suited to an MPC context, i.e., we use the optimal solution of the previous MPC step as an initial guess (warm start) for the next OCP solve. Further, we do not assemble the OCP in each iteration, but rather update only the relevant parts, that is, the initial value $x^n$, the input vector $B(\alpha^n)$ and the reference control $u_{\text{ref},n}$. This allows us to solve \eqref{eq:OCP} with very low computation times in an MPC context, cf.\ Table~\ref{t:comptimes}. The maximal time always occurs in the first MPC-iteration, where no warm start is currently used. The choice and evaluation of a suitable warm start is subject to future work.
\begin{table}[h]
	\centering
	\begin{tabular}{|c||c|c|c|c|c|}
		\hline
		$N$& $2$ & $5$ & $10$ & $15$ & $20$\\\hline
		avg.\ time\,(ms)&0.06 &0.16 &0.29& 0.37&0.41\\
		max.\ time\,(ms)&0.12 &0.34 &0.59& 0.65&0.67\\
		\hline
	\end{tabular}
	\caption{Average and maximal computation time required for solving \eqref{eq:OCP} in an MPC controller. Table taken from \cite{Schaller2022c}.}
	\label{t:comptimes}
\end{table}
\section{Real-time MPC at 250 Hz}
\label{sec:250}
In this part, we present closed-loop MPC results, where the MPC feedback is computed at a repetition rate of 250\,Hz. The corresponding results at the higher repetition rates of 1\,kHz are presented in the subsequent Section~\ref{sec:1}. 

We introduced an appropriate cost functional for \eqref{eq:OCP} that ensures success of the treatment, i.e., the temperature increase should be achieved sufficiently fast, and safety of the treatment, i.e., the peak temperature does not overshoot the temperature bounds. After identifying and comparing in Subsection~\ref{sec:250:sim} suitable configurations for the penalization terms $R_0$, $R_1$ and $R_2$, cf.\ the cost functional \eqref{eq:cost}, we present results of experiments with porcine eye explants in Subsection~\ref{sec:250:exp}.

\subsection{Tuning in simulation}
\label{sec:250:sim}
In this part, we will consider two different simulation plants. First, we will use the reduced order model \eqref{eq:redmodel} that is also considered in the EKF and MPC, so that there is only a model mismatch due to the unknown absorption. Second, we will also consider the full order model \eqref{eq:fullmodel}, leading to a further model-plant mismatch in all components of the model, as the state space dimension is not reduced.
The resulting simulated volume temperature $y_\text{vol}$ is then augmented by noise of one empirical standard deviation, estimated from measurements at the setup, of approximately 0.288\,K \cite{Kleyman2022}.

We will consider two different choices of absorption parameters in our plant:
\begin{itemize}
	\item high absorption of the simulation plant: $\alpha = 1.1$, i.e., the EKF initialized with $\bar{\alpha} = 0.7363$ initially underestimates the parameter,
	\item low absorption of the simulation plant: $\alpha = 0.5$, i.e., the EKF initialized with $\bar{\alpha} = 0.7363$ initially overestimates the parameter.
\end{itemize}

In this part, we will, in addition to different control penalization terms, compare the standard case of state penalization with $R_0\equiv 1$ with the case $R_0(n) = \chi_3(n)$, where
\begin{align*}
\chi_3(n) = \begin{cases}0,\quad n = 1,2,3\\
1, \quad \text{else}
\end{cases}.
\end{align*}
This choice serves to mimick a short identification phase, preventing overly aggressive controls that aim for a fast temperature increase before having a reliable parameter estimation.

We consider a prediction horizon of $N=5$ and compare the following four different choices of the cost functional:
\begin{itemize}
	\item[a)] time invariant state cost, penalization of control offset: $(R_0,R_1,R_2) = (1,5\mathrm{e}4,0)$.
	\item[b)] time variant state cost, penalization of control offset: $(R_0,R_1,R_2) = (\chi_3,5\mathrm{e}1,0)$.
	\item[c)] time invariant state cost, penalization of control variations: $(R_0,R_1,R_2) = (1,0,5\mathrm{e}4)$.
	\item[d)] time variant state cost, penalization of control variations: $(R_0,R_1,R_2) = (\chi_3,0,5\mathrm{e}1)$.
\end{itemize}
These values were found by simulations in 250\,Hz with the full order model \eqref{eq:fullmodel} as plant. They lead to a good performance by means of fast convergence and constraint satisfaction (of the closed-loop  peak temperature) despite the model mismatch due to parameter estimation and model reduction.

\begin{figure}[h]
	\centering
	\includegraphics[width=.8\linewidth]{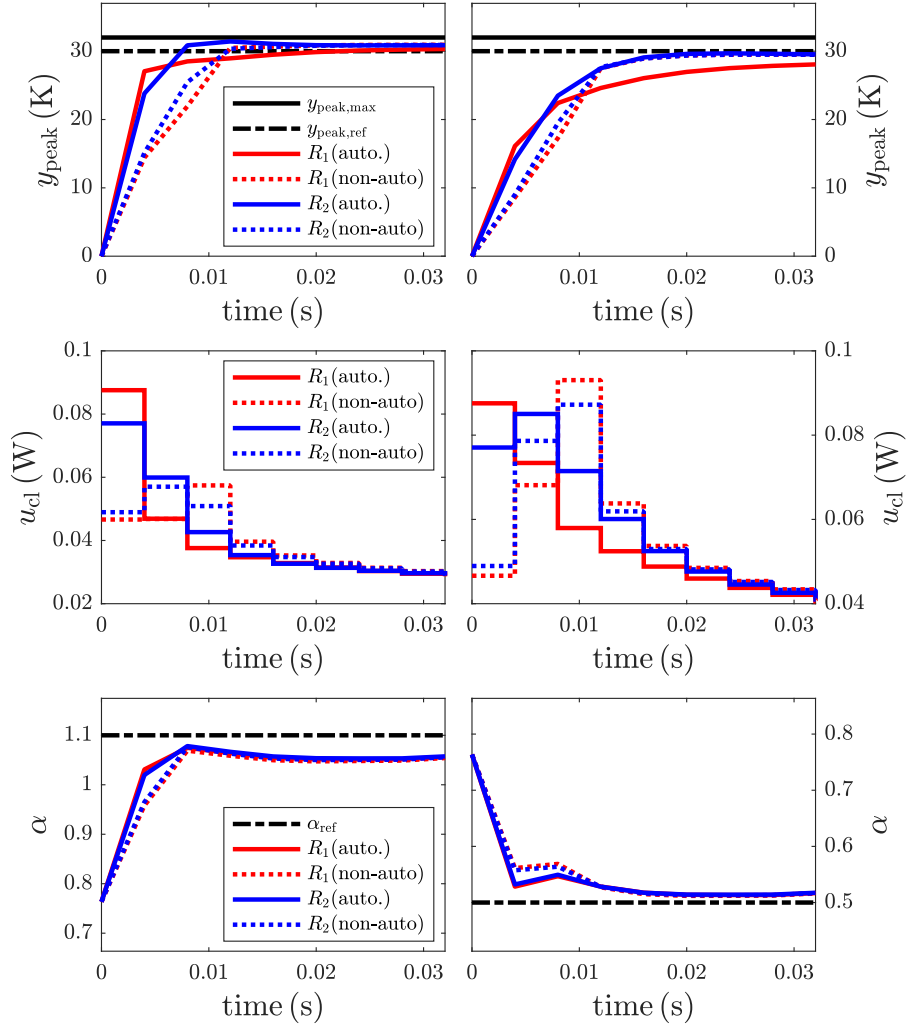}
	\caption{Closed-loop comparison in simulation with full-model plant for $\alpha = 1.1$ (left) and $\alpha = 0.5$ (right). The dotted line corresponds to $R_0=\chi_3$, the solid line corresponds to $R_0\equiv 1$.}
	\label{fig:closed250sim_full}
\end{figure}

We evaluate these different cost functionals for two different scenarios in Fig.~\ref{fig:closed250sim_full}: First, the case of initial underestimation (left), i.e., the initial absorption parameter value for the EKF is smaller than the plant absorption parameter and initial over estimation (right) of the actual parameter, i.e., the initial parameter value is higher than the plant absorption parameter.

We observe that the time-invariant versions ($R_0 = 1$) lead to a faster convergence without significant overshoot in the case of high absorption. The controllers are tuned more aggressively in the first few MPC steps as can be seen in the higher control input $u_\mrm{cl}$ in Fig.~\ref{fig:closed250sim_full}. This is due to the fact that the state cost has a strong influence especially in the first time steps where the difference between the peak temperature and the desired peak temperature is the largest. It follows that $R_1$ and $R_2$, respectively, need to have a higher weight (in comparison to $R_0 = \chi_3$) to prevent constraint violation. On the right hand side of Fig.~\ref{fig:closed250sim_full}, the closed-loop  simulations with a low absorption coefficient are shown. The relation of the control inputs is similar to the high absorption coefficient. However, the time-invariant version with $R_1$ converges slower to the desired peak temperature than both time-varying state costs due to the low absorption and the overestimation at the beginning. This leads to a smaller reference control value $u_\mrm{ref}$ and therefore to a less aggressive controller.

\begin{figure}[h]
	\centering
	\includegraphics[width=.8\linewidth]{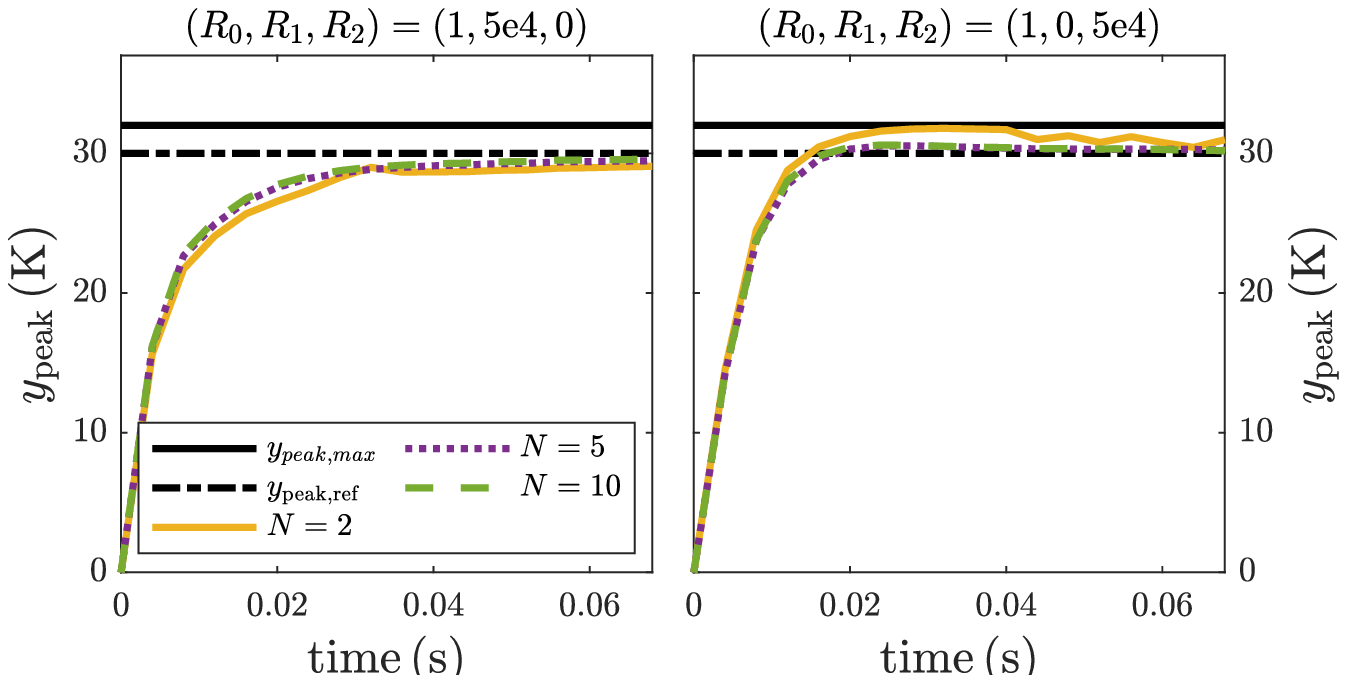}
	\caption{Peak temperature in MPC closed-loop depending on the horizon length for the case $\alpha = 0.5$.}
	\label{fig:horizon250sim}
\end{figure}

\noindent In Fig.~\ref{fig:horizon250sim}, we compare different horizon lengths $N$ for the time-invariant state cost. The case $N=15$ and $N=20$ are almost identical to the case of $N=10$, which is why we do not depict them here. We observe that a horizon length of $N=5$ is suitable for both penalization of the control offset and penalization of the control variations. Higher horizon lengths do not lead to a better performance. Therefore, we choose $N=5$ for the real-time experiments.

\subsection{Real-time experiments with porcine eyes}\label{sec:250:exp}
In this part, we show results by means of experiments with porcine eye explants. As there was no clear advantage of using the non-autonomous cost, we implemented the MPC controller using the autonomous cost functionals with either control or control variation penalization, i.e., $R_0\equiv 1$ in the experimental setup, cf.\ Fig.~\ref{fig:Setup_01}. The non-autonomous case, where $R_0=\chi_3$ in particular in combination with a higher sampling rate of 1\,kHz will be subject to future research.
We compare the resulting closed-loop behaviour for the autonomous cost functionals of the previous section, i.e.,  $(R_0,R_1,R_2) =(1,5e4,0) $ and $(R_0,R_1,R_2) =(1,0,5e4) $ for two different scenarios. 
We emphasize, that our control objective is the peak temperature $y_\mrm{peak}$ which we can only estimate, whereas the volume temperature $y_\mrm{vol}$ is measured and used for state/parameter estimation.

\begin{figure}[h]
	\centering
	\includegraphics[width=.7\columnwidth]{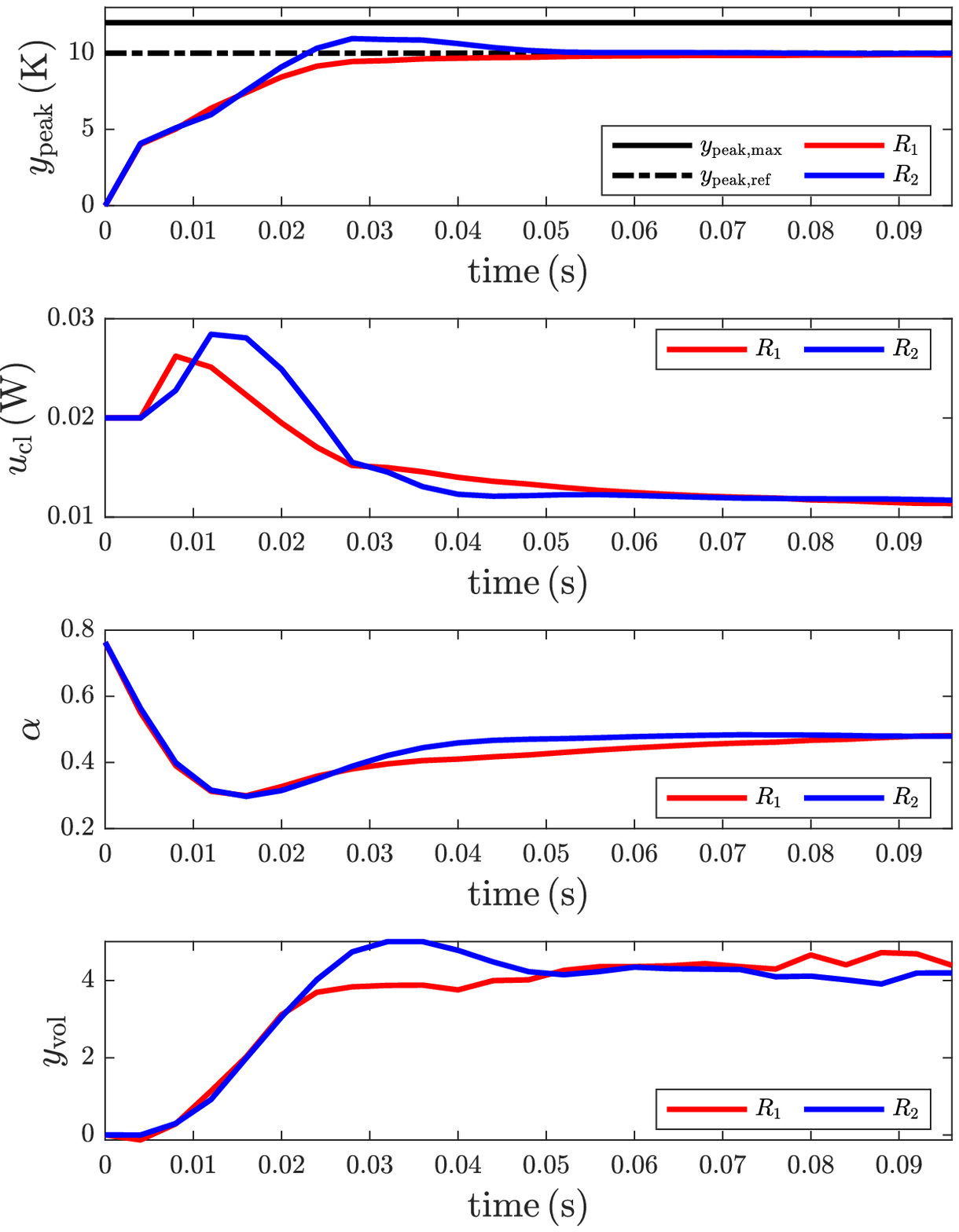}
	\caption{Closed-loop comparison in experiments at the same spot with low treatment temperature.}
	\label{fig:closed0}
\end{figure}

First, in order to compare the MPC closed-loop performance for different cost functionals at the same spot, we consider a low temperature increase $y_\text{peak,ref}= 10\,$K with a maximal temperature increase $y_\text{peak,max}= 12\,$K. This choice prevents from coagulation such that after applying Algorithm~\ref{alg:mpc} with one cost functional choice, we can redo the experiment at the very same spot after changing the cost.
In Fig.~\ref{fig:closed0}, we present results for this low target temperature. The peak temperatures behave similarly and do not exhibit significant overshoot for either choice of the cost functional at the considered treatment spot. The control variation via penalization via $R_2$ slightly overshoots the target temperature of $10\,$K, but does not exceed the peak temperature bound of $12\,$K. The reason for the overshoot of the peak temperature could be the undershoot in the parameter estimation which leads to a more aggressive control strategy. There is no overshoot with the penalization of control offset ($R_1$) as the controller is less aggressive for low parameter values than the one with penalization of the control variations (cf. Fig.~\ref{fig:closed250sim_full}). The MPC closed-loop controls and the identified absorption coefficient are very similar for either of the cost functionals at each spot.

\begin{figure}[h]
	\centering
	\includegraphics[width=.7\columnwidth]{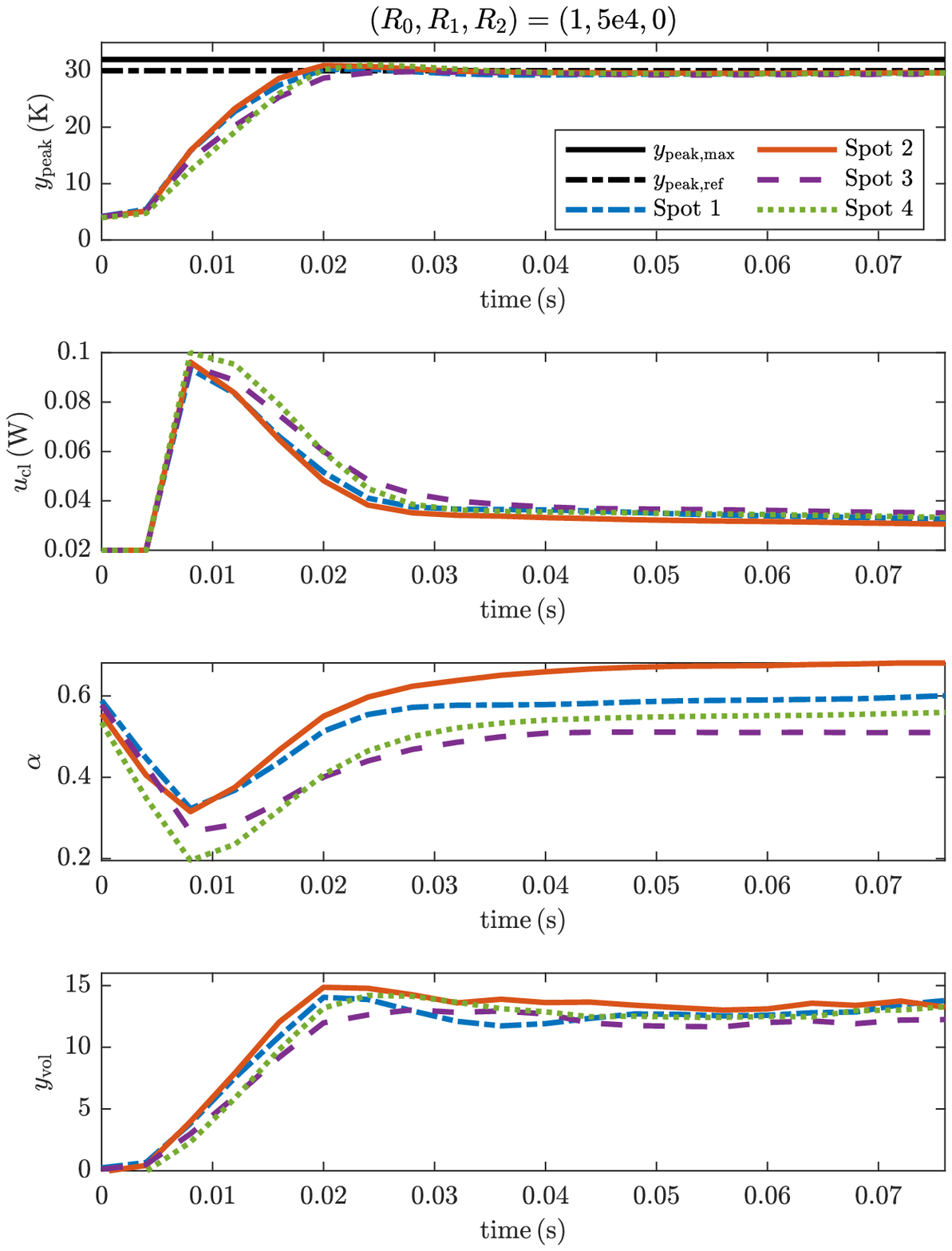}
	\caption{Closed-loop comparison in experiments at different spots with high treatment temperature.}   
	\label{fig:closed3}
\end{figure}

\noindent Second, we consider a targeted temperature increase of $y_\text{peak,ref} = 30\,$K, with a maximal temperature of $y_\text{peak,max} = 32\,$K. As in this case, the tissue is coagulated after one experiment at the same spot, we have to change the treatment spot and thus obtain only one closed-loop trajectory per spot. We note that, here, we obtain the first measurement after applying a constant control of 20\,mW for one time step. 
In Fig.~\ref{fig:closed3}, we consider the target temperature $y_\text{peak,ref}= 30\,$K. Again, we observe an undershoot in the parameter estimation. This undershoot does not affect the closed-loop behaviour when using a direct control penalization by means of $R_1$. Moreover, opposed to the temperature increase of 10\,K in Fig.~\ref{fig:closed0}, we see that for this target temperature of 30\,K, the applied laser power obtained as MPC feedback reaches its bound of $0.1$\,W.
In case of control variation penalization by means of $R_2$, which is not depicted here, the closed-loop peak temperature violated the temperature bounds by approximately $2\,$K.
After increasing the penalization parameter, the choice $R_2 = 8\text{e}5$ guaranteed constraint satisfaction in closed-loop with results looking similar as the closed-loop behavior in Fig.~\ref{fig:closed3}.


\section{Real-time MPC at 1kHz}
\label{sec:1}
In this section, we consider the higher sampling rate of 1\,kHz. At this higher sampling rate, a particular effort has to be made in order to render the complex control software with all its necessary components, i.e., data acquisition, data processing, state and parameter estimation and model predictive control, real-time capable. As we will see in the following, this higher sampling rate enjoys favorable properties, such that it is highly advantageous in terms of state/parameter estimation and feedback-control.
\subsection{Tuning in simulation}

\begin{figure}[h]
	\centering
	\includegraphics[width=.8\columnwidth]{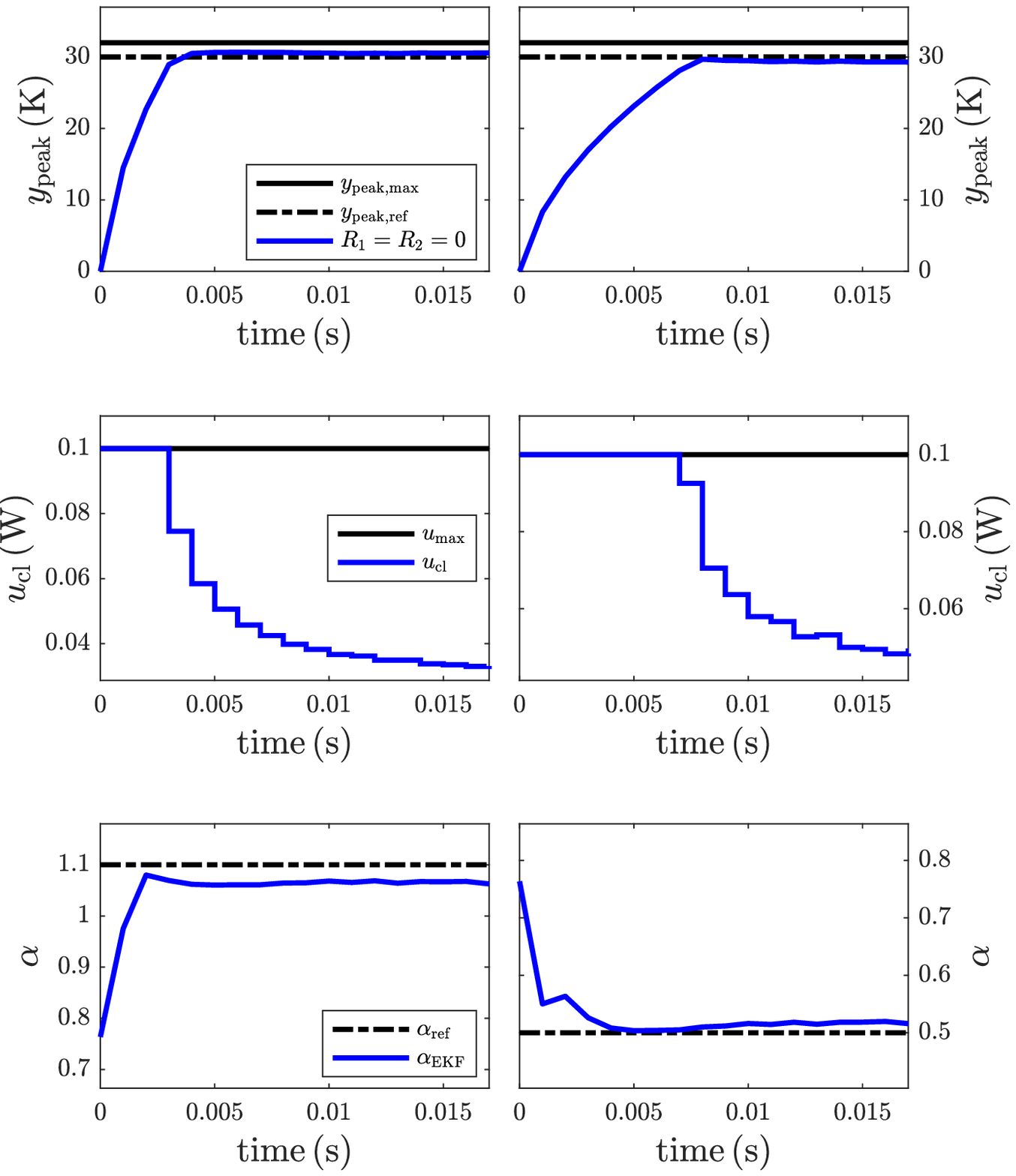}
	\caption{Closed-loop results in simulation with increased sampling rate of 1\,kHz with initial underestimation of the parameter (left) and initial overestimation (right)}
	\label{fig:closed1ksim}
\end{figure}

In Fig.~\ref{fig:closed1ksim}, closed-loop results for an initial underestimation and initial overestimation of the parameter are shown to illustrate differences between the results in 1\,kHz to 250\,Hz. We observe that due to the higher sampling rate, even in the absence of control penalization, i.e., $R_1=R_2=0$, the closed-loop trajectory does not exceed the temperature bounds. This choice not penalizing the controls is also justified in view of our application any used laser power necessary for treatment is satisfactory, as long as it satisfies the bounds $u_\text{max}$. Hence, we do not consider penalization terms here. Further, we can see that input constraints are active, which is to be expected as the controller is most aggressive due to $R_1=R_2=0$. 

\begin{figure}[h]
	\centering
	\includegraphics[width=0.7\columnwidth]{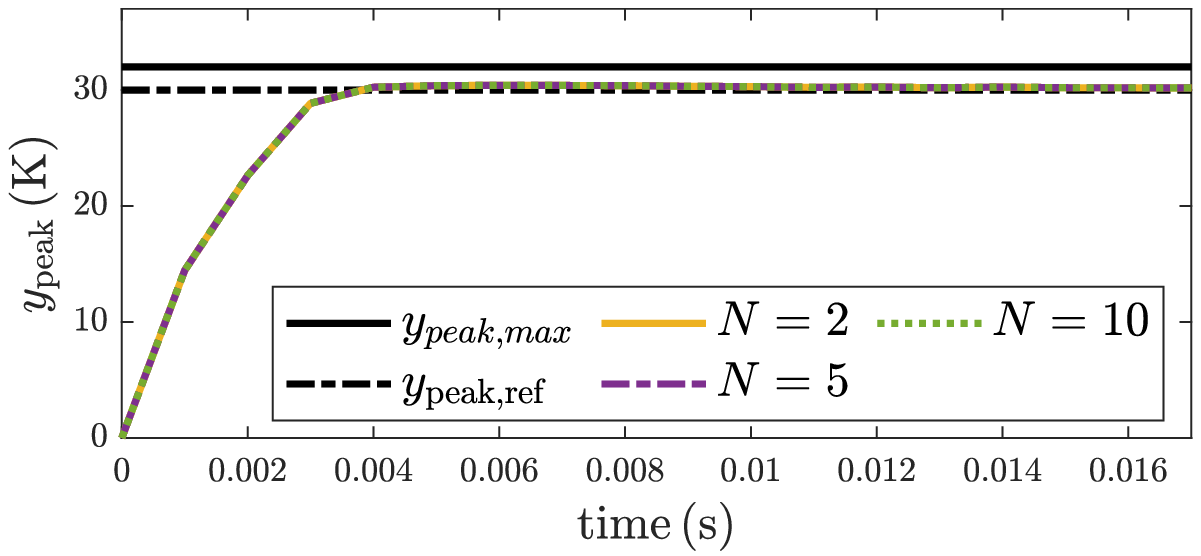}
	\caption{Peak temperature in MPC closed-loop depending on the horizon length for the case $R_1=R_2=0$ with $\alpha = 0.5$.}
	\label{fig:horizon1ksim}
\end{figure}

\noindent Similar to the case of 250\,Hz, we see in Fig.~\ref{fig:horizon1ksim} that the closed-loop performance is mostly independent of the choice of the horizon.
We note that here, a low horizon of $N=2$ is sufficient, as the cost functional term only consists of the state cost, which motivates, independent of the horizon length, a temperature increase whenever the target peak temperature is not yet attained.
\subsection{Real-time experiments with porcine eyes}
\begin{figure}[h]
	\centering
	\includegraphics[width=.7\columnwidth]{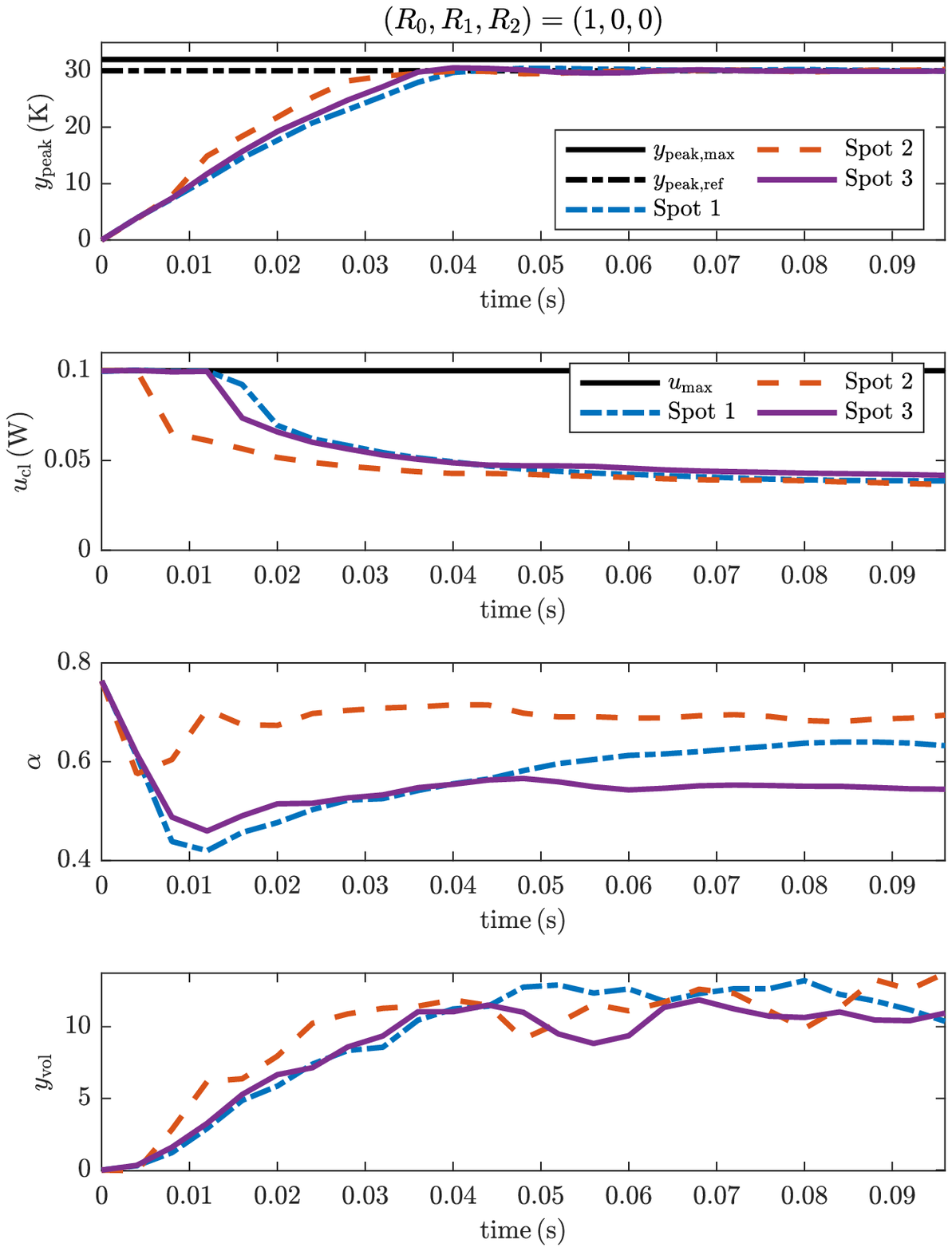}
	\caption{Closed-loop comparison in experiments at different spots with high treatment temperature at 1\,kHz.}   \label{fig:exp1k}
\end{figure}

In Fig.~\ref{fig:exp1k}, we depict the closed-loop behavior for a sampling rate of 1\,kHz for four different spots. In all cases, we observe a convergence to the desired peak temperature increase of 30\,K, without any visible overshoot. At all spots, the laser power constraint of 0.1\,W is active at the beginning of the treatment. For a higher absorption coefficient, the temperature rises faster and, therefore, the laser power constraint is inactive earlier. 

\section{Conclusion and Outlook}\label{sec:conc}

We presented an approach for real-time retinal laser treatment by means of Model Predictive Control. In this context, we designed an appropriate stage cost, for which we evaluated the resulting closed-loop behavior w.r.t.\ convergence properties 
and constraint satisfaction by means of simulation and experimental data in $250$\,Hz and $1$\,kHz. We obtained several configurations of the penalization parameters that result in the desired temperature increase and illustrated that a higher sampling rate leads to better control results with less tuning effort.

Future research considers robust MPC schemes with respect to, e.g., the MOR error, cf.\ \cite{Lorenzetti.2019,Koegel.2015}.

For even faster measurements, e.g., in $10$\,kHz, multi-step MPC schemes~\cite{GrunPann09,GrunPann10} might be an option. Here, the optimal control problem might be solved in~$2$\,kHz, but with variation of the control signal adapted to~$10$\,kHz, see, e.g., \cite{WortRebl14}. Then, the interaction of open- and closed-loop control has to be taken into account~\cite{WortMehr17}, which might be addressed --~if needed~-- by \cite{GrunPalm15,PalmSuar15} or real-time iteration schemes in control~\cite{DiehBock05} and estimation~\cite{WynnVuko14}, see also the recently-proposed advanced step variants of the real-time iteration~\cite{NurkZane19} (and~\cite{NurkZane20} for an analysis of its convergence properties).

\bibliographystyle{abbrv}
\bibliography{references.bib}
\end{document}